\documentclass[aps,prl,twocolumn,footinbib,bibnotes,preprintnumbers,longbibliography]{revtex4-1} 

\usepackage[utf8]{inputenc}
\usepackage{amsmath}
\usepackage{amsfonts}
\usepackage{amssymb}
\usepackage{epsfig}
\usepackage{color}
\usepackage[USenglish]{babel}
\usepackage[normalem]{ulem}

\usepackage{soul,xcolor}
\usepackage{hyperref}
\setstcolor{red}
\usepackage{cleveref}

\allowdisplaybreaks

\crefname{equation}{eq.}{eqs.}

\makeatletter
\newlength{\apb@width}
\newcommand{\autoparbox}[2][c]{\settowidth{\apb@width}{#2}\parbox[#1]{\apb@width}{#2}}
\newcommand{\includegraphicsbox}[2][]{\autoparbox{\includegraphics[#1]{#2}}}
\makeatother

\newcommand{\brk}[1]{(#1)}
\newcommand{\bigbrk}[1]{\bigl(#1\bigr)}

\newcommand{\Complex}{\mathbb{C}}

\newcommand{\polyF}{{\rm{F}}}
\newcommand{\polyU}{{\rm{U}}}
\newcommand{\polyN}{{\rm{N}}}
\newcommand{\Ann}{{\mathbf{A}}}
\newcommand{\Ideal}{\mathcal{I}}
\newcommand{\AnnIdeal}{\cI}

\newcommand{\cD}{\mathcal{D}}
\newcommand{\cI}{\mathcal{I}}
\newcommand{\cJ}{\mathcal{J}}

\newcommand{\soft}[1]{\textsc{#1}}

\def\pd#1{\partial_{#1}}
\newtheorem{conjecture}{Conjecture}

\begin{document}
\newpage
\pagenumbering{arabic}

\title{Differential Space of Feynman Integrals:  
Annihilators and $\cD$-module }

\newcommand{\padova}{
    Dipartimento di Fisica e Astronomia, Universit\`a degli Studi di Padova
    e INFN, Sezione di Padova,
    Via Marzolo 8, I-35131 Padova, Italy.
}

\newcommand{\bologna}{
    Dipartimento di Fisica e Astronomia, Universit\`a di Bologna
    e INFN, Sezione di Bologna,
    via Irnerio 46, I-40126 Bologna, Italy.
}

\newcommand{\liverpool}{Department of Mathematical Sciences, University of Liverpool, Liverpool L69 3BX, 
United Kingdom
}

\author{Vsevolod Chestnov,}
\affiliation{
    Dipartimento di Fisica e Astronomia, Universit\`a di Bologna
    e INFN, Sezione di Bologna,
    via Irnerio 46, I-40126 Bologna, Italy.
}

\author{Wojciech Flieger}
\author{Pierpaolo Mastrolia,}
\affiliation{Dipartimento di Fisica e Astronomia, Universit\`a degli Studi di Padova, Via Marzolo 8, I-35131 Padova, Italy}
\affiliation{INFN, Sezione di Padova, Via Marzolo 8, I-35131 Padova, Italy}

\author{Saiei-Jaeyeong Matsubara-Heo}
\affiliation{Graduate School of Information Sciences, Tohoku University, Aoba Aramaki-aza 6-3-09, Aoba Ward, Sendai 980-8579, Japan}

\author{Nobuki Takayama}
\affiliation{Department of Mathematics, Kobe University, 1-1, Rokkodai, Nada-ku, Kobe 657-8501, Japan}

\author{William J. Torres~Bobadilla}
\affiliation{\liverpool}

\begin{abstract}
We present a novel algorithm for constructing differential operators with respect to external variables that annihilate Feynman-like integrals and give rise to the associated $\cD$-modules, based on Griffiths–Dwork reduction. 
By leveraging the Macaulay matrix method, we derive corresponding relations among partial differential operators, including systems of Pfaffian equations and Picard-Fuchs operators.
Our computational approach is applicable to twisted period integrals in projective coordinates, and we showcase its application to Feynman graphs and Witten diagrams. The method yields annihilators and their algebraic relations for generic regulator values, explicitly avoiding contributions from surface terms. In the cases examined, we observe that the holonomic rank of the $\cD$-modules coincides with the dimension of the corresponding de Rham co-homology groups, indicating an equivalence relation between them, which we propose as a conjecture.
 
\end{abstract}

\setcounter{tocdepth}{2}
\maketitle
\setcounter{page}{1}

\section{Introduction}

In perturbative field theory, Feynman integrals play a central role for the evaluation of scattering amplitudes, correlation functions, and related physical observables, across various domains of physics, including particle physics, general relativity, and cosmology, both in quantum and classical contexts.
These integrals are part of a broader class known as twisted period integrals, which also encompass Aomoto-Gelfand hypergeometric functions and Euler integrals.
The study of special functions, necessary to evaluate Feynman integrals, has become an active area of research, relating differential and algebraic geometry, topology, and number theory to Feynman calculus, and 
fostering a rich interplay between mathematics and physics.

Scattering amplitudes, Green functions, and Feynman integrals satisfy differential equations, such as the Callan-Symanzik equation, Lorentz invariance identities, and Picard-Fuchs equations, or are annihilated by operators that, according to the considered theory, may reflect their symmetries and homogeneity with respect to external parameters.
Therefore, developing methods to derive the differential operators that annihilate twisted period integrals is crucial for classifying the special functions representing these integrals, providing an additional tool alongside direct integration.
Within the standard approach, differential equations for Feynman  integrals~\cite{Kotikov:1990kg, Kotikov:1991pm,Remiddi:1997ny,Gehrmann:1999as,Argeri:2007up,Henn:2013pwa,Argeri:2014qva,Papadopoulos:2014lla} are derived by means of integration-by-parts (IBP) identities  \cite{Tkachov:1981wb,Chetyrkin:1981qh,Laporta:2000dsw}, in momentum space representation,
and more recently by vector space projection making use of intersection numbers for twisted de Rham co-homology~\cite{Mastrolia:2018uzb,Frellesvig:2019kgj,Frellesvig:2019uqt}, in parametric representation~\footnote{
    First studies of differential equations for Feynman integrals, earlier than the introduction of the dimensional regularization, also made use of the parametric representation~\cite{Regge:1965,Giffon:1969se,Barucchi:1973}, and more recently in~\cite{Bern:1993kr}
}{\nocite{Regge:1965,Giffon:1969se,Barucchi:1973,Bern:1993kr}}.

Interesting studies pointed to the role of the Griffiths-Dwork pole reduction method~\cite{Griffiths_1969,Dwork1962,Dwork1964} for deriving Picard-Fuchs equations of Feynman integrals in parametric representations \cite{Muller-Stach:2012tgj,Zeng:2017ipr,Bonisch:2021yfw,Lairez:2022zkj,Artico:2023jrc,delaCruz:2024xit},
building the homogeneous term of the differential operator (which annihilates the so-called {\it maximal-cut} of the integral) and treating the inhomogeneous term, referred to as the {\it surface term}, coming from sub-graphs, as a contribution to be independently evaluated.

Recent studies on $\cD$-module techniques for Feynman and generalized Euler integrals \cite{Chestnov:2022alh,Chestnov:2023kww,Henn:2023tbo, Pfister:2025} showed that identities among elements of the module can be used to derive relations among differential operators which parallel the IBP relations among integrals.

\section{Methodology}

In this {\it Letter}, we propose a novel approach for constructing differential operators w.r.t. external variables that annihilate Feynman integrals in $d$ dimensions, hereafter referred to as {\it annihilators}. 
These operators form the so-called {\it annihilating ideal} in the space of all differential operators $\cD$ with polynomial coefficients, and can be used to identify the corresponding $\cD$-module~\footnote{
    Annihilators of a given function $\pi\brk{s}$ form an ideal: if $\Ann$ annihilates $\pi\brk{s}$, that is $\Ann \, \pi\brk{s} = 0$, then left multiplication with any other differential operator will do so too and thus belong to the ideal as well.
    The associated $\cD$-module is the space of all differential operators $\cD$ modulo those that annihilate $\pi\brk{s}$.
},
providing a natural algebraic framework for analyzing Feynman integrals and allowing for the construction of the associated Pfaffian systems and Picard-Fuchs (PF) differential equations.

Our approach is inspired by the study of generalized Euler integrals, for which Gelfand, Kapranov, and Zelevinsky (GKZ) proposed an algorithm to derive generators of the annihilating ideal directly from the parametric integral representation \cite{Gelfand1989}.
This ideal gives rise to the GKZ $\cD$-module, which was proven to be {\it equivalent} to the twisted de Rham co-homology group~\cite{Gelfand:1990}%
\footnote{
    For a list of applications of the GKZ approach to Feynman integrals, see~\cite{%
    Nasrollahpoursamami:2016,%
    delaCruz:2019skx,%
    Klausen:2021yrt,%
    Klausen:2019hrg,%
    Klausen:2023gui,%
    Kalmykov:2020cqz,%
    Tellander:2021xdz,%
    Pal:2021llg,%
    Pal:2023kgu,%
    Ananthanarayan:2022ntm,%
    Feng:2022kgh,%
    Feng:2022ude,%
    Feng:2019bdx,%
    Zhang:2023fil,%
    walther2022feynman,%
    Dlapa:2023cvx,%
    Agostini:2022cgv,%
    Munch:2022ouq,%
    Klemm:2019dbm,%
    Bonisch:2020qmm,%
    delaCruz:2024ssb,%
    Grimm:2024tbg%
}.
}\nocite{%
    Nasrollahpoursamami:2016,%
    delaCruz:2019skx,%
    Klausen:2021yrt,%
    Klausen:2019hrg,%
    Klausen:2023gui,%
    Kalmykov:2020cqz,%
    Tellander:2021xdz,%
    Pal:2021llg,%
    Pal:2023kgu,%
    Ananthanarayan:2022ntm,%
    Feng:2022kgh,%
    Feng:2022ude,%
    Feng:2019bdx,%
    Zhang:2023fil,%
    walther2022feynman,%
    Dlapa:2023cvx,%
    Agostini:2022cgv,%
    Munch:2022ouq,%
    Klemm:2019dbm,%
    Bonisch:2020qmm,%
    delaCruz:2024ssb,%
    Grimm:2024tbg%
}.

However, in the case of Feynman integrals, considered as {\it restrictions} of Euler integrals \cite{Chestnov:2022alh,Chestnov:2023kww}, a similar algorithm to deduce $\cD$-module presentations directly from the integral representations is not available.
In this communication, we elaborate and extend the results of \cite{Golubeva:1973,Golubeva:1978}, and use the twisted version of Griffiths-Dwork (GD) reduction to build annihilators w.r.t. external variables~\footnote{Annihilators of Feynman integrals w.r.t. integration variables were earlier considered in \cite{Bitoun:2017nre}}, which turn out to have a remarkably simple structure, 
being formed just by the combination of two differential operators of adjacent order, say $p$ and $p-1$.  
We use them to construct 
$\cD$-modules, and by means of the Macaulay matrix method \cite{Chestnov:2022alh}, we identify a basis of the $\cD$-module,
known as {\it standard monomials}, whose size identifies to the {\it holonomic rank}. Relations among differential operators, resulting from solving the Macaulay matrix, yield the derivation of Pfaffian systems for standard monomials, and PF operators in the absence of surface terms, in alternative to the IBP derivation to the (systems of) differential equations obeyed by Feynman integrals.

Earlier studies considered the remarkable correspondence of Euler characteristic $\chi$
with the holonomic rank of $\cD$-modules, and with 
the dimension of the de Rham co-homology group, 
corresponding to the number of master integrals (MI),
both for Feynman integrals \cite{Bitoun:2017nre,Lee:2013hzt,Mastrolia:2018uzb,Frellesvig:2019uqt}, and for generalized Euler integrals of GKZ type 
\cite{Agostini:2022cgv,Matsubara:2025mht}. 
In the latter case, $\chi$ is known to match the mixed volume of Newton polyhedra \cite{Gelfand1989,Adolphson-1994,SaitoSturmfelsTakayama::book}.
In general, $\chi$ is given by the maximal likelihood degree of a toric variety \cite{10.2307/40067993,Matsubara:2025mht}.

The applications of our novel approach to paradigmatic cases, including special functions, Feynman integrals, and Witten integrals, point toward a new potential {\it correspondence} between the $\cD$-module and the twisted de Rham co-homology group, indicating a non-trivial extension of the GKZ theorem to restricted classes of integrals. 
Within our framework, the dimension of the co-homology group, hence the number of MI, is related to the holonomic rank of the $\cD$-module of differential operators w.r.t. external variables and to Euler characteristic, and thus to the order of the PF  differential equations satisfied by generic twisted period integrals.

\section{Setup}

We consider dimensionally regulated (scalar) Feynman integrals corresponding to $\ell$-loop diagrams.
In parametric representation, they can be interpreted as twisted period integrals, written as:
\begin{align}
\label{eq:feynman_parametrization}
\pi(s)=\int_{\Gamma} \Omega \ ,   
\quad {\rm with } \quad 
 \Omega \equiv u \, \varphi \ , 
 \end{align}
where the {\it twist} $u$ and the differential form $\varphi$ are defined as,
\begin{align}
& u \equiv \frac{ \polyU(\alpha)^{\kappa}}{ \polyF(\alpha, s)^{\eta}} \, 
\quad {\rm and } \quad
\varphi \equiv 
\prod_{i=1}^n \alpha_i^{\nu_i-1} \, 
\mu \,, 
\end{align}
and where 
$\alpha = (\alpha_1, \ldots , \alpha_{n} )$ 
and $s= (s_1, \ldots , s_{m})$ 
are the sets of integration and of the external variables, respectively.
The integration measure is defined as
$
\mu \equiv \sum _{i=1}^{n} (-1)^{i-1}\alpha_{i} \, d\alpha_{1} \wedge \cdots \wedge \widehat{d\alpha_{i}} \wedge \cdots \wedge d\alpha_{n} \,,
$
and the domain of integration (twisted cycle) $\Gamma= \lbrace (\alpha_{1}, \hdots , \alpha_{n}) \vert
\sum_{i=1}^{n}\alpha_{i} =1 \,, \alpha_{i} \geq 0 \rbrace$.
$\polyU$ and $ \polyF$ are the first and second Symanzik polynomials:
they are homogeneous polynomials of degree $\ell$ and $\ell+1$ respectively, $\ell \in \mathbb{Z_{+}}$; 
$ \polyF$ is also homogeneous in $s$ of degree $1$. 
The exponents
$\kappa = \sum_{i=1}^n \nu_i - (\ell+1)d/2$ 
and $\eta = \sum_{i=1}^n \nu_i - \ell d/2$, 
depend on $\ell$, the space-time dimension $d$, and on the number of integration variables $n$.
Since $d \in \Complex$ is generically a non-integer number, also $\kappa$ and $\eta$ are non-integers, making the integrand $\Omega$ a multi-valued function.

\section{Griffiths-Dwork reduction and Annihilators}

GD reduction \cite{Griffiths_1969,Dwork1962,Dwork1964} was conceived to reduce the order of the pole of rational differential forms. Elaborating on the work of \cite{Golubeva:1973,Golubeva:1978}, 
we hereby employ the {\it twisted} version of GD reduction \cite{delaCruz:2024xit}, 
loaded with special boundary conditions, 
to build elements $\Ann_p$ belonging to the annihilating ideal
$ {\cal I} = \{ \Ann_p \ | \ \Ann_p \pi(s)=0 \}$, namely formed by 
differential operators w.r.t. the external $s$ variables, 
that annihilate $\pi(s)$.

In this study, we consider Feynman integrals with unit denominators' exponents,
$\nu_i = 1 \ (i=1,\ldots,n)$, since, we experience they are sufficient to derive $\cD$-modules~\footnote{
Applications to integrals with ${\mathbb N} \ni \nu_i > 1$ can be derived along the same lines.
}.
Our proposal consists in building the annihilators $\Ann_p$, 
\begin{equation}
     \Ann_p \equiv {\bf D}_p + {\bf D}_{p-1} \ , 
    \label{eq:annihilators}
\end{equation}
in terms of 
two generic differential operators of order $p$ and $p-1$, respectively introduced through the ansatz,
\begin{align}
& {\bf D}_{p} = \sum_{I , \vert I \vert = p} c_{I} \, \partial_s^{I} \ ,
\label{eq:diff_p_deg}
\end{align}
where, using the multi-index notation,
$I= \lbrace i_{1}, \hdots, i_{m} \rbrace$, 
$\partial_s^I \equiv \partial^{(I)}/\partial s^I$ 
stands for the $\vert I \vert $-th order partial derivative w.r.t. the {\it external} variables, 
and $c_I$ 
are generic coefficients to be later determined.
The direct applications of
${\bf D}_p$ and ${\bf D}_{p-1}$ to $\pi(s)$ in \cref{eq:feynman_parametrization}
generate integrals of the type,
\begin{align}
& {\bf D}_{p} \, \pi(s) = 
\int_{\Gamma} \frac{ \polyN_p}{ \polyF(\alpha, s)^{p}} \Omega \,, \label{eq:op_p}
\end{align} 
where $\polyN_p$ is polynomial in $\alpha_i$. Therefore, in our construction, we build ${\bf A}_p$ as
\begin{align}
\label{eq:annihilators2}
    {\bf A}_p \, \pi(s)
    = ({\bf D}_{p} + {\bf D}_{p-1}) \, \pi(s) = 
    \int_{\Gamma} d \gamma
    = \int_{\partial \Gamma} \! \gamma
    = 0 \ ,
\end{align}
where we look for differential form $\gamma$ that vanishes at the boundary of the integration domain.

According to the twisted GD reduction, we may look for such a $\gamma$ in the following form
\begin{align}\label{eq:twisted_gamma}
\begin{split}
\gamma &\equiv \frac{1}{ \polyF^{k-1}} \sum_{i<j} (-1)^{i+j} \left( \alpha_{i}\lambda_{j} - \alpha_{j}\lambda_{i} \right)
u\\
&\times
d\alpha_{1} \wedge \hdots \widehat{d\alpha_{i}} \wedge \hdots \wedge \widehat{d\alpha_{j}} \hdots \wedge d\alpha_{n} \,.   
\end{split}
\end{align}
for $k \in {\mathbb N}$.
The corresponding expression of $d\gamma$ reads as:
\begin{align}
\label{eq:twisted_griffiths}
d \gamma = (\eta + k-1)
\frac{
\sum_{i=1}^{n} \lambda_{i} \, \partial_i \polyF
}{\polyF^{k}} \Omega 
- \frac{
\sum_{i=1}^{n} 
\nabla_i \, \lambda_i
}{\polyF^{k-1}}
\Omega \,, 
\end{align}
with the {\it twisted covariant derivative} $\nabla_i$, defined as,
\begin{eqnarray}
      \nabla_i \equiv \partial_i 
    + \omega_i \wedge \ , \quad {\rm with} \quad 
    \omega_i \equiv \partial_i {\rm log}(\polyU^\kappa) \ ,
\end{eqnarray}
where we used the shorthand notation 
$\partial_i \equiv \partial/\partial \alpha_i$, for the partial derivative w.r.t. the {\it integration} variables.
Thus, the requirements of~\cref{eq:annihilators2} can be satisfied with the following \textit{three} conditions:
\begin{itemize}
\item[A)] Boundary condition:
\begin{align}
\lambda_{i} \Big|_{\alpha_{i}=0} = 0 \,;
\label{eq:cond:boundary}
\end{align}
\item[B)] Jacobian ideal membership condition:
\begin{align}
    \polyN_p = \sum_{i=1}^{n} \lambda_{i} \, 
    \partial_i \polyF
    \,,
\label{eq:cond:jacobian}
\end{align}
implying that $\polyN_p \in
\langle 
\partial_1 \polyF, \ldots, 
\partial_n \polyF
\rangle$, being the Jacobian ideal of $\polyF$; 
\item[C)] Twisted covariant derivative  condition:
\begin{align} 
\polyN_{p-1} = \sum_{i=1}^{n} 
\nabla_i \, \lambda_i \,.
\label{eq:cond:gaussmanin}
\end{align}
\end{itemize}

The annihilators $\Ann_p$ can be finally built
by imposing these constraints on the functions $\lambda_i=\lambda_i(c_I,s,\alpha)$, that are polynomials in $\alpha$, which depend linearly on the coefficients $c_I=c_I(s)$. Let us remark that the conditions 
(B) and (C) were also considered in \cite{delaCruz:2024xit}, while 
the determination of $\lambda_i$, also obeying the condition (A), is a key feature of our algorithm, that avoids the appearance of inhomogeneous surface terms.

\section{$\cD$-modules}
Let us consider the $m$-th rational Weyl algebra~\footnote{
    See, e.g., \cite{dojo, ot-2001, SST} for a review of $\cD$-modules theory and related computational techniques. Relations among the rational Weyl algebra $R_m$, the polynomial Weyl algebra $\cD$, and their sheaf-theoretic counterparts are also discussed in these references.
}\nocite{dojo, ot-2001, SST}
\begin{align}
R_m=\Complex(s_1, \ldots, s_m)\langle \pd{1}, \ldots, \pd{m} \rangle\,,
\end{align}
that is a ring of differential operators
with rational function coefficients.
It is an associative ring subject to the relations
\begin{align}
[\pd{i},\pd{j}]=0\,,  \quad  
\pd{i} c(s) = c(s) \pd{i} + {\partial c(s) \over \partial s_i} \,,\label{eq:D relation}
\end{align}
where $c(s)$ is a rational function of $s=(s_1, \ldots, s_m)$.

A left ideal $\Ideal$ of the rational Weyl algebra $R_m$ can be regarded as a system of linear partial differential equations.
The \textit{holonomic rank} is defined as the dimension of quotient $R_m/\Ideal$
when viewed as a vector space over the field of rational functions $\Complex(s)=\Complex(s_1, \ldots, s_m)$
\begin{equation}
\label{eq:holonomic_rank}
    r(\Ideal) \equiv {\rm dim}_{\Complex(s)} \, R_m/\Ideal
    \,.
\end{equation}
In this study, we work exclusively with the rational Weyl algebra $R_m$.
To simplify the exposition, and with slight abuse of terminology, we refer to $R_m/\Ideal$ as a $\cD$-module \footnote{
The theory of $\cD$-modules is developed for the polynomial Weyl algebra
$
    \cD=\Complex[s,\dots,s_m]\langle\pd{1},\dots,\pd{m}\rangle\subsetneq R_m
$
with the same commutation relation \labelcref{eq:D relation} as in the rational Weyl algebra $R_m$.
The correspondence between the two setups comes from the observation, that for any $\Ideal$ in $R_m$ there always exists a \brk{non-unique} ideal $\cJ$ in $\cD$.
The holonomic rank~\labelcref{eq:holonomic_rank} agrees with the dimension of the holomorphic solutions of the system $\cJ$ at a generic point, see, e.g., \cite[6.1--6.3]{dojo} and a generalization of this result by Kashiwara in derived categories \cite{kashiwara-1970}.
}.
Accordingly, when we speak of a basis of a $\cD$-module, we mean a basis of the corresponding $R_m$-module when regarded as a vector space over the field $\Complex\brk{s}$.

\begin{conjecture}
Under genericity conditions on $\polyU$, $\polyF$, $\kappa$, $\eta$
and when $p$ is sufficiently large,
the holonomic rank $r$ of the left ideal of the rational
Weyl algebra $R_m$
generated by
${\bf D}_i+{\bf D}_{i-1}$, $i \leq p$
is  $r=|\chi(V)| \,$ 
where $\chi(V)$ is the Euler characteristic of the zero set $V$ of $\polyU(\alpha)\polyF(\alpha,s)$
in the $\alpha$-space $\{ \alpha_1\cdots\alpha_n\neq 0\}$ with generic $s$-parameters values.
\end{conjecture}
Let us observe that by the inclusion-exclusion property of Euler characteristics~\cite{ALUFFI_2019,Bitoun:2017nre,Frellesvig:2019uqt,Mizera:2019ose,Matsubara:2025mht}, 
our Conjecture implies,
\begin{align}
\label{eq:eulerXi}
r & = |\chi(V)| 
= |n-\chi(V_{E}({f}))|  = {\rm dim}(H^{n-1}_{\rm dR})
 \ ,    
\end{align} 
 where 
$V_E(f)$ is the vanishing locus of 
$f(\alpha_1,\dots, \alpha_{n}) \equiv
 \alpha_1 \dots \alpha_{n}  
 \polyU(\alpha) \polyF(\alpha,s)$ for generic $s$,
in the projective space $\mathbb{P}^{n-1}$.
It namely means that 
 the rank $r$ of the $\cD$-module coincides with the dimension of $H^{n-1}_{\rm dR}$ \cite{Frellesvig:2019uqt,Mizera:2019ose,Matsubara:2025mht}, 
 denoting either the {\it twisted co-homology group} \cite{cho1995} or equivalently the {\it relative twisted co-homology group} \cite{matsumoto2024relative} --- since the relation holds in both cases
\footnote{In \cref{eq:eulerXi} the following relation is employed:
$
\chi(\{\alpha_1\cdots\alpha_n\neq 0\}\setminus V)  = 
$
$
    \chi(\{\alpha_1\cdots\alpha_n\neq 0\}) 
    -\chi(V) 
$
$    = -\chi(V) 
$
$    = \chi(\mathbb{P}^{n-1})-\chi(V_E(f)) 
$
$    = n-\chi(V_E(f)) \ , 
$
with 
$\chi(\{\alpha_1\cdots\alpha_n\neq 0\})=0$
and 
$\chi(\mathbb{P}^{n-1})=n$.
For further details on the correspondence between the Euler characteristic and the dimension of the associated twisted co-homology group for the integrand, 
see e.g. \cite{Matsubara:2025mht} and 
references therein.}.

Dimensionally regularized $\ell$-loop Feynman integrals are, by definition, homogeneous in the external variables 
$s$~\footnote{
    From the momentum space representation, assuming standard quadratic denominators, the degree of homogeneity is $\ell d/2 - \sum_i \nu_i$, being $\nu_i$ the denominator powers.
}. Therefore, there always exist a differential operator ${\bf A}_E = {\bf D}_1+{\bf D}_0$ that annihilates $\pi(s)$, and takes the form 
\begin{equation} \label{eq:euler_homog}
{\bf A}_E = e_1 s_1 \pd{1} + \cdots + e_m s_m \pd{m}-1 \ ,
\end{equation}
where $e_i \in \Complex$.
The key distinction between our approach and conventional methods dealing with holonomic systems 
\cite{chyzak:tel-01069831,Lairez_2015,brochet2025} is that we express generators of the ideal as the sum of two homogeneous differential operators ${\bf D}_i$ and ${\bf D}_{i-1}\,$.
The existence of the annihilator \labelcref{eq:euler_homog} contributes to this property.
Although we do not know yet the sufficient conditions under which this conjecture holds, and we have only proven it for certain toy models, the conjecture turns out to be valid in several physically interesting cases, as we will see later.

\section{Algorithm}

Our computational strategy has three goals: 
{\it i)} building the differential operators $\Ann_p$ that generate the annihilating ideal 
${\cal I}$; 
{\it ii)} using them within the Macaulay method \cite{Chestnov:2022alh}, 
to identify a basis of 
standard (Std) monomials that generate the $\cD$-module;
{\it iii)} and deriving the decomposition formulas of the partial differential operators in terms of Std monomials,
as well as the PF equation  
satisfied by $\pi(s)$.

\noindent
{\bf Annihilators.}
The conditions (A) and (B) can be simultaneously fulfilled by choosing 
$\lambda_i$ as \begin{equation}\label{eq:lambdaansatze}
    \lambda_i = {q_i \over q_0} \alpha_i 
\end{equation}
in terms of the polynomials 
$q_0 = q_0(s,c_I)$ and 
$q_i=q_i(s,c_I,\alpha)$, 
with $i=1,\ldots,n \,,$
that, upon inserting \cref{eq:lambdaansatze}
in \cref{eq:cond:jacobian},
can be found as solutions of the syzygy equation,
\begin{align}\label{eq:rescale_syz}
\sum_{i=1}^{n}
q_{i} \, \theta_i \polyF = 
q_{0} \, \polyN_{p} \,.
\end{align}
for the ideal 
$ 
\langle 
\theta_{1} \polyF , \ldots , 
\theta_{n} \polyF, \polyN_{p} \rangle
\,,$
with $\theta_i \equiv \alpha_{i} \partial_{i}$
being the Euler operators.

After imposing (A) and (B), the condition (C), namely eq.  \eqref{eq:cond:gaussmanin},
turns into a linear systems for the coefficients $c_{I}$, whose solution
yields  
parametric expressions of the differential operators ${\bf D}_p$ and ${\bf D}_{p-1}$ depending on the independent coefficients $c_{I,{\rm ind}}$. 
The annihilators $\Ann_p$ are finally built
 by {\it choosing} the values $c_{I,{\rm ind}}$: operationally, we found that setting all but one of the $c_{I,\text{ind}}$ to a non-zero value, at once, and going through the entire set of the independent coefficients produces the sufficient set of representatives of $\AnnIdeal$~\footnote{
The solutions $q_i$ in \eqref{eq:lambdaansatze} of the inhomogeneous syzygy equation \eqref{eq:rescale_syz} are defined up to solutions $q_{i,H}$ of the homogeneous syzygy equation 
$\sum_{i} q_{i,H} \, \theta_i {\polyF}=0$. In our algorithm, this freedom is exploited for the determination of $\lambda_i$, satisfying the A, B, C conditions, to build annihilators also in the case of functions with non-isolated singularities \cite{Lairez_2015}.}.

\noindent
{\bf Standard monomials.}
After constructing the set of annihilators $\AnnIdeal$, the independent generators of  the $\cD$-module can be derived using the Macaulay matrix method. This approach relies on the solution of a linear system of equations, whose matrix $M$ is built from coefficients of monomials in the partial derivatives of the operators $\Ann_{p} \in \AnnIdeal$ 
\begin{align}
M = (\partial^{I} \Ann_{p}) = 0 \,, \quad \vert I \vert = 0, \ldots , l \,,   
\label{eq:macaulay_matrix}
\end{align}
where $l$ is a chosen maximal degree of the derivatives,
{\it i.e.} 
$\partial^I = \partial_{s_1}^{i_1} \cdots \partial_{s_m}^{i_m}$ 
with $i_1 + \ldots i_m = |I|$.
The columns of this matrix are labeled by the partial derivative monomials appearing in the operators~\labelcref{eq:macaulay_matrix}, which can be organized w.r.t. the degree-lexicographic order.
Row reduction of the system then produces the set of independent monomials, so-called {\it standard monomials}, defined as,
\begin{align}
{\rm Std} = \lbrace  \partial^{k} \rbrace \,.    
\end{align}
The size of ${\rm Std}$ is equal to the holonomic rank $r$ in~\cref{eq:holonomic_rank}, 
therefore our conjecture implies,  
    \begin{align}
        r = |{\rm Std}| = {\rm dim}H_{\rm dR}^{n-1}\,.
    \end{align}
\noindent
    In other words, the system $M = 0$ is a linear system of differential operators which parallels the IBP system of equations for the Feynman integrals, where the Std's play the role of master integrals \cite{Chestnov:2022alh}.  

\noindent
{\bf Pfaffian systems and Picard-Fuchs operators.}
By choosing the appropriate monomial ordering, 
   the Macaulay system can be solved to 
identify a basis of Std monomials, and derive the Pfaffian system of equations they obey: the latter naturally emerge from the decomposition of partial derivatives acting on Std's modulo the generating ideal.
    Moreover,
    for sufficiently large $l$ values in \cref{eq:macaulay_matrix}, the independent annihilators can be combined into operators that involve  derivatives w.r.t. only one variable, which can be identified with the PF operators. The PF operators can be then easily transformed into a Pfaffian system in the basis of derivatives of $\pi(s)$.

\section{Applications}\label{sec:Applications}
We apply our method to derive the annihilating differential operators and build the corresponding $\cD$-modules in 
paradigmatic cases of special functions, like the ${}_2F_1$ and ${}_3F_2$
hypergeometric integrals, Feynman integrals and Witten graphs, in dimensional regularization.
In the considered cases, Euler homogeneity annihilator $\Ann_{E}$ is among the annihilators $\Ann_{p}$ that our algorithm produces.
Annihilators are used to generate the Macaulay system, which we solve 
to identify a basis of 
{\rm Std} monomials, derive the relations they obey, and the {\it complete} PF operator $\Ann_{\rm PF}$ w.r.t. one of the external variables. 
In all applications,
 the holonomic rank $r$
 is found to be equivalent to 
 ${\rm dim}(H_{\rm dR}^{n-1})$, providing evidence of the  conjectured correspondence between the 
 $\cD$-module and the twisted de Rham co-homology group. 
 Additionally, in the case of Feynman integrals,  the singular locus of the generated $\cD$-modules is verified to agree with Landau singularities of the first and second type,
  and the PF operators are verified to agree with those built by means of IBP identities, 
  constituting non-trivial tests of the generated $\cD$-modules~\footnote{
    Our algorithm is implemented in {\sc Mathematica} \cite{Mathematica}. 
    The solution of syzygy equations, the holonomic rank and the singular locus of the generated $\cD$-modules 
    are computed within the computer algebra system {\sc Singular} \cite{DGPS}.
    Euler characteristic of the zero locus of affine varieties are computed with {\sc Macaulay2} \cite{M2}.
    For {Gr\"obner} basis computations in Weyl algebra, we use \soft{Risa/Asir} \cite{url-asir}.
    Linear systems and Macaulay matrix equations are solved with the help of {\sc FiniteFlow}~\cite{Peraro:2019svx}.
    Factorization of Picard-Fuchs equation is carried out in {\sc Maple} \cite{maple}
}. \\

\noindent
{\it Notation}.
In the following, for each application, we define the Symanzik polynomials $\polyU$ and 
$\polyF$, and their exponents $\kappa$ and $\eta$. 
Lengthier generators of the annihilating ideals $\AnnIdeal$
and PF operators are given in implicit form, as a combination of differential operators multiplied by generic polynomial coefficients or rational functions, say $p=p(d,s)$ and $r=r(d,s)$: while just their structure is relevant to highlight the main results of this study, interested readers may find their complete expressions in the {\it Supplementary Material}. \\

\subsection{ a) One-loop massless 4-point integral}

\begin{figure}[h!]
\includegraphicsbox[]{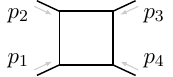}
\caption{ One-loop massless 4-point graph. Kinematic variables: 
$p_{1}^{2} = \hdots = p_{4}^{2} = 0 \,, s= 2 p_{1} \cdot p_{2} \,, t = 2 p_{2} \cdot p_{3}$.
\label{fig:massless_box}
}
\end{figure}

Our first example is the massless integral shown in~\cref{fig:massless_box},
which was also discussed in~\cite{delaCruz:2024xit}.

\noindent
$\bullet$ Symanzik polynomials:
\begin{align}
\label{eq:massless_box_integrand}
\polyU = \sum_{i=1}^{4} \alpha_{i} \,, 
\quad {\rm and } \quad
 \polyF =  \alpha_{1} \alpha_{3} s + \alpha_{2} \alpha_{4} t \,,
 \end{align}
with exponents $\kappa = 4-d$ and  
$\eta = 4 - d/2 $. \\

\noindent
$\bullet$ Generators of $\AnnIdeal$:
\begin{align}
\begin{split}
&\Ann_{E} = s \partial_{s} + t \partial_{t} + (4-d/2) \,, \\
& \Ann_{3} = p_{3,0} \, 
 \partial_{s}^{3}
 +p_{2,1} \, 
 \partial_{s}^{2}\partial_{t} 
 +p_{1,2} \, 
 \partial_{s} \partial_{t}^{2}
+ p_{0,2} \,
\partial_{t}^{2} \,.
\label{eq:ann_massless_box}
\end{split}
\end{align}
\noindent
$\bullet$ $\cD$-module rank $r=3$, 
and singular locus:
\begin{align}
{\rm sing}(\Ideal) = st(s+t) \,.    
\end{align}

\noindent
$\bullet$ Picard-Fuchs operator:
\begin{align}
\Ann_{\rm PF} &= 
\left(\partial_{s}+r_{1}\right)\cdot\left(\partial_{s}+r_{2}\right)\cdot\left(\partial_{s}+r_{3}\right)\,.
\label{eq:pf_massless_box}
\end{align}

\subsection{ b) One-loop one-mass 4-point integral}

\begin{figure}[h!]
\includegraphicsbox[]{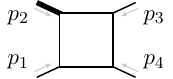}
\caption{One-loop 1-mass 4-point graph.
Kinematic variables:
$p_{1}^{2} = p_{3}^{2} = p_{4}^{2} = 0 \,, p_{2}^{2} = m^2 \,,$ 
$ 2 p_{1} \cdot p_{2} = (s-m^2) \,, 
\ 2 p_{2} \cdot p_{3} = (t-m^2)  \,.$ 
\label{fig:one_mass_box}
}
\end{figure}

Next, we add one external mass to the four-point graph as depicted
in~\cref{fig:one_mass_box}. 

\noindent
$\bullet$ Symanzik polynomials:
\begin{align}
 {\polyU} = \sum_{i=1}^{4} \alpha_{i} \,, 
 \ {\rm and} \ 
{\polyF} =  \alpha_{1} \alpha_{3} s + \alpha_{2} \alpha_{4} t + \alpha_{2}\alpha_{3} m^2 \,.
\end{align}
with exponents 
$ \kappa = 4-d \,,$ and 
$ \eta = 4 - d/2 $. \\

\noindent
$\bullet$ Generators of $\AnnIdeal$:
\begin{align}
 & \Ann_{E} = s\partial_{s} + t \partial_{t} + m^2 \partial_{m^2} 
 + \left(4- d/2\right) 
 \,, \notag\\
& \Ann_{2,1} = s \partial_{s}\partial_{t} + (m^2-t)\partial_{t}\partial_{m^2} + \partial_{t} -\partial_{m^2} \,, \\
& \Ann_{2,2} = -s \partial_{s}^{2} +t \partial_{t}\partial_{m^2} + m^2 \partial_{m^2}^{2} + 
 (4-d/2)
 \left( \partial_{m^2} - \partial_{s} \right) \, .
 \notag
\end{align}

\noindent
$\bullet$ $\cD$-module rank $r=4$, and singular locus:
\begin{align}
{\rm sing}(\Ideal) = stm^{2}(s-m^2)(t-m^2)(s+t-m^2) \,.    
\end{align}

\noindent
$\bullet$ Picard-Fuchs operator:
\begin{equation}
    \Ann_{\rm PF} = 
\left(\partial_{s}+r_{1}\right)\cdot\left(\partial_{s}+r_{2}\right)^2\cdot\left(\partial_{s}+r_{3}\right)\,.
\label{eq:pf_one_mass_box}
\end{equation}

\subsection{c) Three-loop equal mass 2-point integral}

\begin{figure}[h!]
\centering
\includegraphicsbox[]{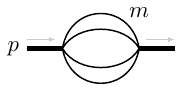}
\caption{
Three-loop banana graph. 
Kinematic variables: 
$p^{2} = s$, and identical-mass propagators, $m_{1}^{2}= \ldots = m_{4}^{2} = m^2$. 
\label{fig:3_loop_banana}
}
\end{figure}

Our higher perturbative order example is the self-energy graph illustrated
in~\cref{fig:3_loop_banana} earlier studied in \cite{Pogel:2022yat}.

\noindent
$\bullet$ Symanzik polynomials are:
\begin{align}
\begin{split}
&\polyU = \alpha_{1}\alpha_{2}\alpha_{3} + \alpha_{1}\alpha_{2}\alpha_{4} + \alpha_{1}\alpha_{3}\alpha_{4} +\alpha_{2}\alpha_{3}\alpha_{4} \,, \\
&\polyF = \alpha_{1} \alpha_{2} \alpha_{3} \alpha_{4} s-U \sum_{i=1}^{4} \alpha_{i} m^{2} \,,
\end{split}
\end{align}
with exponents
$\kappa = 4 - 2 d$ and $\eta = 4 - (3/2) d$.\\

\noindent     
$\bullet$ Generators of $\AnnIdeal$:
\begin{align}
\begin{split}
&\Ann_{E} = s\partial_{s} + m^{2}\partial_{m^2} 
+ (4 - (3/2)d) \,, \\
&\Ann_4 =
    p_{4,0} \, \partial_{s}^{4} 
    + p_{0,3} \, \partial_{m^{2}}^{3} 
    + p_{1,2} \, \partial_{s}\partial_{m^{2}}^{2} 
     \\
    & \qquad \quad
    + p_{2,1} \, \partial_{s}^{2}\partial_{m^{2}} 
    + p_{3,0} \, \partial_{s}^{3} \,.
\label{eq:ann_3_loop_banana}
\end{split}
\end{align}

\noindent
$\bullet$ $\cD$-module rank $r=4$, and singular locus:
\begin{align}
{\rm sing}(\Ideal) = sm^{2}(s-4m^2)(s-16m^2) \,.    
\end{align}

\noindent
$\bullet$ Picard-Fuchs operator:
\begin{align}
\Ann_{\rm PF} &= 
    \left(\partial_{s}+r_{1}\right)\cdot\left(\partial_{s}^{3}+r_{2}\partial_{s}^{2}+r_{3}\partial_{s}+r_{4}\right)\,.
    \label{eq:pf_3l_banana}
\end{align}

\subsection{d) One-loop  Witten integral in $\text{AdS}_{4}$}

\begin{figure}[h!]
    \centering
\includegraphicsbox[]{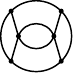}
\caption{One-loop $s$-channel Witten diagram in AdS$_4$. }
    \label{fig:witten_diag_1L}
\end{figure}

Finally, we consider the integral corresponding to the $s$-channel one-loop graph in~\cref{fig:witten_diag_1L} studied in \cite{Heckelbacher:2022fbx}, hereby homogenized w.r.t. external variables.

\noindent
$\bullet$ Symanzik polynomials:
\begin{align}
\polyU_{1}= & (\alpha_{2}+\alpha_{3}+\alpha_{4}+\alpha_{5})\alpha_{1}+(\alpha_{2}+\alpha_{3}+\alpha_{4})\alpha_{5}\,, \\
\polyU_{2}= & \alpha_{1}\,,\quad\polyU_{3}=\alpha_{3}\,,\quad\polyU_{4}=\alpha_{5}\,,\nonumber\\
\polyF= & \alpha_{2}(\alpha_{3}\alpha_{5}+\alpha_{1}(\alpha_{3}+\alpha_{5}))\zeta_{0}^{2}+\alpha_{2}\alpha_{4}(\alpha_{1}+\alpha_{5})\zeta\bar{\zeta}\nonumber \\
 & +\alpha_{4}(\alpha_{3}\alpha_{5}+\alpha_{1}(\alpha_{3}+\alpha_{5}))(\zeta_{0}-\zeta)(\zeta_{0}-\bar{\zeta})\,,\nonumber 
\end{align}
with exponents
$\kappa_{1} = -1-\epsilon$, $\kappa_{2} = -2 \epsilon$, $\kappa_{3} = -1 - 2\epsilon$, $\kappa_{4} = 1$ and $\eta = 2\epsilon -1$. 

\noindent
$\bullet$ Generators of $\AnnIdeal$:
\begin{align}
 & \Ann_{E}=\zeta_{0}\partial_{\zeta_{0}}+\zeta\partial_{\zeta}++\bar{\zeta}\partial_{\bar{\zeta}}+(2-4\epsilon)\,,\label{eq:1l_witten_op}\\
 & \Ann_{2,1}=p_{2,0,0}\partial_{\zeta}^{2}+p_{0,2,0}\partial_{\bar{\zeta}}^{2}+p_{0,0,2}\partial_{\zeta_{0}}^{2}+p_{1,1,0}\partial_{\zeta}\partial_{\bar{\zeta}}\nonumber \\
 & +p_{1,0,1}\partial_{\zeta}\partial_{\zeta_{0}}+p_{0,1,1}\partial_{\bar{\zeta}}\partial_{\zeta_{0}}+p_{1,0,0}\partial_{\zeta}+p_{0,1,0}\partial_{\bar{\zeta}}+p_{0,0,1}\partial_{\zeta_{0}}\,,\nonumber \\
 & \Ann_{2,2}=q_{2,0,0}\partial_{\zeta}^{2}+q_{0,2,0}\partial_{\bar{\zeta}}^{2}+q_{0,0,2}\partial_{\zeta_{0}}^{2}+q_{1,1,0}\partial_{\zeta}\partial_{\bar{\zeta}}+\nonumber \\
 & q_{1,0,1}\partial_{\zeta}\partial_{\zeta_{0}}+q_{0,1,1}\partial_{\bar{\zeta}}\partial_{\zeta_{0}}+q_{1,0,0}\partial_{\zeta}+q_{0,1,0}\partial_{\bar{\zeta}}+q_{0,0,1}\partial_{\zeta_{0}}\,.\nonumber 
\end{align}

\noindent
$\bullet$ $\cD$-module rank $r=4$, and singular locus 
(affine variables , i.e. $\zeta_{0} \to 1$):
\begin{align}
{\rm sing}(\Ideal) = \zeta \bar{\zeta} (\zeta-1) (\bar{\zeta}-1) (\zeta-\bar{\zeta}) \,.    
\end{align}

The rank of the $\cD$-module is $r=|\chi(V)|$, where $\chi(V)$ is computed with eq.\eqref{eq:eulerXi}, for $n=5$, 
and 
$f(\alpha_1,\alpha_2, \alpha_3,\alpha_{4}, \alpha_{5}) = 
 \alpha_2  \alpha_{4}  
 \polyU_1 \polyU_2 \polyU_3 \polyU_4 \, \polyF$, noticing that $\polyU_2 \polyU_3 \polyU_4 = \alpha_1 \alpha_3 \alpha_5$, 
 so that $f$ is as in eq.\eqref{eq:eulerXi}.

\noindent
$\bullet$ Picard-Fuchs operator (affine variables , i.e. $\zeta_{0} \to 1$):
\begin{align}\label{eq:1l_witten_pf}
\Ann_{\rm PF}
=
\left(\partial_{\zeta}+r_{1}\right)\cdot\left(\partial_{\zeta}+r_{2}\right)\cdot\left(\partial_{\zeta}^{2}+r_{3}\partial_{\zeta}+r_{4}\right)\,.
\end{align}

The PF operator, exact in $d=4-2\epsilon$, for this diagram, is derived here for the first time.
We observe that in the limit $\epsilon \to 0$, 
$\Ann_{\rm PF}$  factorizes into the product of four linear terms, as $(\partial_\zeta+r_1)\cdot\hdots \cdot(\partial_\zeta+r_4)$\, 
therefore implying that the solution can be written in terms of polylogarithms, confirming the direct integration result, up to the leading order in $\epsilon$ (c.f., eq.~(4.18) in~\cite{Heckelbacher:2022fbx}). Moreover, we verified the annihilation of this function up to this order with the above operators.

\section{Conclusion} 
We studied the differential space of  Feynman integrals in parametric representation and proposed an algorithm based on Griffiths-Dwork reduction to build generators of the annihilating ideal with respect to the external variables, used for a presentation of the $\cD$-module, having a remarkably simple form.
Through the Macaulay matrix method, we derived a minimal set of independent generators 
of the $\cD$-module,
and found that its holonomic rank corresponds to the dimension of the twisted de Rham co-homology groups, hence to the number of master integrals. 
The identity between the dimensions of the space of differential operators 
and of the vector space of integrals induced us to consider the existence of an {\it equivalence relation} among them, which would extend the proposition of Gelfand, Kapranov and Zelevinski's theorem to the (wider) class of restricted integral cases.

Our study on the structure of the {\it differential space} of Feynman-like integrals naturally complements  the {\it vector space} structure which emerged from the application of intersection theory for twisted de Rham co-homology~\cite{Mastrolia:2018uzb,Frellesvig:2019kgj,Frellesvig:2019uqt}.
The proposed method, grounded in generic conditions, is expected to extend well beyond Feynman integrals. It offers a systematic approach for deriving annihilators, Pfaffian systems, and Picard-Fuchs operators in a broad range of contexts—including scattering amplitudes, correlation functions, and other special functions associated with holonomic systems that admit a twisted period representation. As such, it holds potential significance for both mathematical and physical applications. 
Combining the ideas of the annihilating ideal  as proposed in this study together with the {\it shift-operator algebra} of \cite{Pfister:2025,Matsubara-Heo:2025lrq} may represent a very interesting future direction.

\section*{Acknowledgments}
We aknowledge G. Brunello, S. Cacciatori, L. De La Cruz, C. Fevola, G. Fontana, J. Henn, M. Mandal, A. Massidda, T. Peraro, A.L. Sattelberger, P. Vanhove, S. Weinzierl, for discussions and comments on the manuscript, and 
G. Brunello, for valuable discussions on syzygies computation.
The work of V.C. was supported by the European Research Council (ERC) under the
European Union's Horizon Europe research and innovation program grant agreement
101040760, \textit{High-precision multi-leg Higgs and top physics with finite
fields} (ERC Starting Grant \emph{FFHiggsTop}).
W.F. and P.M. acknowledge the support of the INFN research initiative {\it Amplitudes}. W.F. would like to acknowledge the support of the Polish National Science Centre (NCN) under grant 2023/50/A/ST2/00224.  
The work of W.J.T. was supported by the Leverhulme Trust, LIP-2021-01.
W.J.T. acknowledges hospitality of the Universit\`a degli Studi di Padova and INFN sezione di Padova where parts of this work were completed.

\section*{Supplemental Material}\label{sec:appendix}

\section*{Twisted Griffiths-Dwork formula}
For completeness, we hereby report the derivation of 
formula (\ref{eq:twisted_griffiths}), earlier  considered in \cite{delaCruz:2024xit}.
By taking the total differential w.r.t. integration variables of \cref{eq:twisted_gamma} we get
\begin{align}\label{eq:twisted_d_gamma}
d \gamma = (k-1)\frac{\sum_{i=1}^{n} \lambda_{i} u \, \partial_{i} \polyF }{\polyF^{k}}\mu -  \frac{\sum_{i=1}^{n} \partial_{i} (\lambda_{i} u) }{\polyF^{k-1}}\mu \,,   
\end{align}
where $\partial_{i} = \frac{\partial}{\partial \alpha_{i}}$.
Let us write explicitly the second term on the right hand side
\begin{align}
&\frac{\sum_{i=1}^{n} \partial_{i} (\lambda_{i} u) }{\polyF^{k-1}}\mu =  \frac{1}{\polyF^{k-1}} \sum_{i=1}^{n} \left[  \frac{\partial}{\partial \alpha_{i}} \left(  \lambda_{i} u \right)    \right]\mu =  \notag \\
& \frac{1}{\polyF^{k-1}} \sum_{i=1}^{n} \left[  u \frac{\partial \lambda_{i}}{\partial \alpha_{i}}   +  \kappa \lambda_{i}\frac{\polyU^{\kappa-1}}{\polyF^{\eta}} \frac{\partial \polyU}{\partial \alpha_{i}} -\eta \lambda_{i}\frac{\polyU^{\kappa}}{\polyF^{\eta+1}} \frac{\partial \polyF}{\partial \alpha_{i}}  \right]\mu \\
&= \frac{1}{\polyF^{k-1}} \sum_{i=1}^{n} \left[ \frac{\partial \lambda_{i}}{\partial \alpha_{i}}   +  \kappa \lambda_{i} \frac{\partial \log(\polyU)}{\partial \alpha_{i}} -\eta \lambda_{i} \frac{\partial \log(\polyF)}{\partial \alpha_{i}}  \right] u\mu \notag \,.
\end{align}
Plugging it back to \cref{eq:twisted_d_gamma}, we get get \cref{eq:twisted_griffiths}, i.e.,
\begin{align}
d \gamma = (\eta + k-1)
\frac{
\sum_{i=1}^{n} \lambda_{i} \, \partial_i \polyF
}{\polyF^{k}} \Omega 
- \frac{
\sum_{i=1}^{n} 
\nabla_i \, \lambda_i
}{\polyF^{k-1}}
\Omega \,.
\end{align}

\section*{Hypergeometric Integrals}
In this section, we showcase the application of our algorithm to derive the generators ${\bf A}_p$ of the $\cD$-modules of two classical hypergeometric functions, and the related Picard-Fuchs equation as elements of the $\cD$-modules.

\subsection{ ${}_2F_1$ {\bf integral}}

\noindent
$\bullet$ Defining polynomials (in projective coordinates):
\begin{align}
    & \polyU_1 = \alpha_1 \brk{\alpha_2 - \alpha_1}
    \>, \quad 
    \polyU_2 = \alpha_2 
    \>, \\
    & \polyF = x_2 \alpha_2 - x_1 \alpha_1
    \>.
\end{align}
with exponents $\kappa_1 = \eta$, $\kappa_2 = -3 \eta - 2$.

\noindent
$\bullet$ Generators of $\AnnIdeal$:
\begin{align}
    \Ann_E &= x_1 \> \partial_{x_1} + x_2 \> \partial_{x_2} - \eta
    \label{eq:2F1_euler}
    \\
    \Ann_{2,1} &=
    x_1 \brk{x_2 - x_1} \> \partial_{x_1}^2
    + \bigbrk{
        - \brk{1 - \eta} x_1
        + 2 \brk{1 + \eta} x_2
    } \> \partial_{x_1}
    \nonumber
    \\
    &\quad
    + \brk{1 + \eta} x_2 \> \partial_{x_2}
    \>,
    \\
    \Ann_{2,2} &=
    \brk{x_2 - x_1} \> \partial_{x_1} \partial_{x_2}
    - \brk{1 + 3 \eta} \> \partial_{x_1}
    - \brk{1 + \eta} \> \partial_{x_2}
    \>,
    \\
    \Ann_{2,3} &=
    x_2 \brk{x_2 - x_1} \> \partial_{x_2}^2
    + \brk{1 + 3 \eta} x_1 \> \partial_{x_1}
    \nonumber
    \\
    &\quad
    + \bigbrk{
        2 \eta \> x_1
        + \brk{1 - \eta} x_2
    } \> \partial_{x_2}
    \>.
\end{align}

\noindent
$\bullet$ $\cD$-module rank: 2. 

\noindent
$\bullet$
Picard-Fuchs equation w.r.t. the ratio $t = x_1 / x_2$:
\begin{align}
    \Ann_{\rm PF} = 
        t \brk{1 - t} \> \partial_t^2
        + 2 \brk{1 + \eta - t} \> \partial_t
        + \eta \brk{1 + \eta} \ .
    \label{eq:2F1_deq}
\end{align}
in agreement with the well known result in the mathematical literature.

\subsection{ ${}_3F_2$ {\bf integral}}

\noindent
$\bullet$ Defining polynomials (in projective coordinates):
\begin{align}
    &\polyU_1 = \alpha_1 \> \alpha_2 \> \brk{\alpha_1 - \alpha_2} \> \brk{\alpha_1 - \alpha_3} \,,
    \quad
    \polyU_2 = \alpha_3 \,,
    \\
    &\polyF = x_2 \alpha_2 - x_1 \alpha_3 \,,
\end{align}
with exponents $\kappa_1 = \eta$, $\kappa_2 = -3 - 5 \eta$.

\noindent
$\bullet$ Generators of $\AnnIdeal$:
The space of differential operators is spanned by the Euler
operator~\labelcref{eq:2F1_euler} and four degree-3 annihilators.

\noindent
$\bullet$  $\cD$-module rank: 3.

\noindent
$\bullet$
Picard-Fuchs equation w.r.t. the ratio $t = x_1 / x_2$:
\begin{align}
    \Ann_{{\rm PF}} = 
    &t^2 \brk{1 - t} \> \partial_t^3
    + 3 \eta \, t \brk{-2 + 3 t} \> \partial_t^2
+ \eta \, \bigbrk{
        2\brk{1 + 4 \eta}
    \nonumber
    \\
&        - \brk{5 + 23 \eta} \, t
    } \> \partial_t
    + \eta \, \brk{1 + 3 \eta} \brk{2 + 5 \eta}
    \>.
\end{align}
in agreement with the well known result in the mathematical literature.

\section*{Generators of $\AnnIdeal$}

In this section, interested readers may find the complete expression of the annihilators $\Ann_p$ generated by our algorithm, which were given in implicit form in Section {\it Applications} of the {\it Letter}. \\

\subsection{a) One-loop massless 4-point integral} 
Full expression of \cref{eq:ann_massless_box}
\begin{align}
 \Ann_{3} &= p_{3,0} \, 
 \partial_{s}^{3}
 +p_{2,1} \, 
 \partial_{s}^{2}\partial_{t} 
 +p_{1,2} \, 
 \partial_{s} \partial_{t}^{2}
+ p_{0,2} \,
\partial_{t}^{2} \,,
\end{align}
where
\begin{align}
p_{3,0} &= - 4 s^2 \,, \notag \\
p_{2,1} &= s  ((d-8)^2 s+2 (26-3 d) t) \,, \notag \\
p_{1,2} &= - t (2(28-3 d) s+(d-8)^2 t) \,, \notag \\
p_{0,2} & = 2(d-12) t \,. \notag
\end{align}
Full expression of \cref{eq:pf_massless_box} 
\begin{align}
\Ann_{\rm PF} &= 
\left(\partial_{s}+r_{1}\right)\cdot\left(\partial_{s}+r_{2}\right)\cdot\left(\partial_{s}+r_{3}\right)\,,
\end{align}
where, 
\begin{align}
r_{1}=&-\frac{d-10}{2s}-\frac{1}{u}\,,
\notag\\
r_{2}=&\frac{1}{s}-\frac{1}{u}\,,
\notag\\
r_{3}=&-\frac{1}{2}\left(\frac{d-4}{u}+\frac{d-6}{s}\right)\,,
\end{align}
and $u=-s-t$. \\

\subsection{{b) One-loop one-mass 4-point integral}}
\noindent Full form of \cref{eq:pf_one_mass_box}
\begin{equation}
    \Ann_{\rm PF} = 
\left(\partial_{s}+r_{1}\right)\cdot\left(\partial_{s}+r_{2}\right)^2\cdot\left(\partial_{s}+r_{3}\right)\,,
\end{equation}
where, 
\begin{align}
r_{1}= & -\frac{d-10}{2s}-\frac{1}{u}-\frac{1}{m^{2}-s}\,,\nonumber \\
r_{2}= & \frac{1}{s}-\frac{1}{u}-\frac{1}{m^{2}-s}\,,\nonumber \\
r_{3}= & -\frac{1}{2}\left(\frac{d-4}{u}+\frac{d-6}{s}\right)\,,
\end{align}
and $u=m^{2}-s-t$.

\subsection{{c) Three-loop equal mass 2-point integral}}
Full form of \cref{eq:ann_3_loop_banana}
\begin{align}
    \Ann_{4} &=
    p_{4,0} \, \partial_{s}^{4} 
    + p_{0,3} \, \partial_{m^{2}}^{3} 
    + p_{1,2} \, \partial_{s}\partial_{m^{2}}^{2} 
    \notag \\
    & \qquad 
    + p_{2,1} \, \partial_{s}^{2}\partial_{m^{2}} 
    + p_{3,0} \, \partial_{s}^{3} \,,
\end{align}
where
\begin{align}
    p_{4,0} & = 6 \left(3 d^2-22 d+40\right) s^2 \left(s^2-20 m^{2} s+64 m^{4}\right)
    \,, \notag \\
    p_{0,3} & = - 4 \left(d^2-7 d+12\right) m^{6}
    \,,\notag \\
    p_{1,2} & = - 6 m^{4} \Big(\left(d^3-15 d^2+70 d-104\right) s+ \notag\\
& \qquad \qquad \qquad 2 \left(d^3+d^2-62 d+160\right) m^{2}\Big)
    \,,\notag \\
    p_{2,1} & = - m^{2} \Big(\left(-21 d^3+266 d^2-1100 d+1488\right) s^2+ 
    \notag\\
&4 \left(27 d^3-508 d^2+2656 d-4200\right) m^{2} s+ \notag\\
&64 \left(3 d^3-22 d^2+4 d+120\right) m^{4}\Big)
    \,, \notag \\
    p_{3,0} & = -s \Big(3 \left(9 d^3-108 d^2+428 d-560\right) s^2- \notag \\
&4 \left(111 d^3-1424 d^2+5888 d-7920\right) m^{2} s+ \notag\\
&64 \left(3 d^3-112 d^2+664 d-1080\right) m^{4}\Big)
    \,. \notag 
\end{align}
Full form of \cref{eq:pf_3l_banana}
\begin{align}
\Ann_{\rm PF} &= 
    \left(\partial_{s}+r_{1}\right)\cdot\left(\partial_{s}^{3}+r_{2}\partial_{s}^{2}+r_{3}\partial_{s}+r_{4}\right)\,,
\end{align}
where, 
\begin{align}
    r_{1}&=\frac{2}{s}-\frac{1}{4m^{2}-s}-\frac{1}{16m^{2}-s}\,,
    \notag\\
    r_{2}&=\frac{3}{2}\left(\frac{d-3}{4m^{2}-s}+\frac{d-3}{16m^{2}-s}+\frac{2}{s}\right)\,,
    \notag\\
    r_{3}&=\frac{1}{s}\left(\frac{2(d-3)}{4m^{2}-s}-\frac{(3d-10)(d-3)}{16m^{2}-s}-\frac{(d-4)d}{4s}\right)\,,
    \notag\\
    r_{4}&=\frac{(d-3)(3d-8)}{8s^{2}}\left(\frac{(3d-10)}{16m^{2}-s}-\frac{(d-2)}{4m^{2}-s}\right)\,.
\end{align}

\subsection{{d) One-loop Witten integral in $\text{AdS}_{4}$ }}

\noindent Full expression of \cref{eq:1l_witten_op}
\begin{align}
\begin{split}
&\Ann_{2,1} = p_{2,0,0} \partial_{\zeta}^{2} + p_{0,2,0} \partial_{\bar{\zeta}}^{2} + p_{0,0,2} \partial_{\zeta_{0}}^{2} + p_{1,1,0} \partial_{\zeta}\partial_{\bar{\zeta}} + \\
&p_{1,0,1} \partial_{\zeta}\partial_{\zeta_{0}} + p_{0,1,1} \partial_{\bar{\zeta}}\partial_{\zeta_{0}}  + p_{1,0,0} \partial_{\zeta} + p_{0,1,0} \partial_{\bar{\zeta}} + p_{0,0,1} \partial_{\zeta_{0}} \,, \\
&\Ann_{2,2} = q_{2,0,0} \partial_{\zeta}^{2} + q_{0,2,0} \partial_{\bar{\zeta}}^{2} + q_{0,0,2} \partial_{\zeta_{0}}^{2} + q_{1,1,0} \partial_{\zeta}\partial_{\bar{\zeta}} + \\
&q_{1,0,1} \partial_{\zeta}\partial_{\zeta_{0}} + q_{0,1,1} \partial_{\bar{\zeta}}\partial_{\zeta_{0}}  + q_{1,0,0} \partial_{\zeta} + q_{0,1,0} \partial_{\bar{\zeta}} + q_{0,0,1} \partial_{\zeta_{0}} \,,
\end{split}
\end{align}
where
\begin{align*}
p_{2,0,0} &= \zeta \big(\zeta^2 (\bar{\zeta} (19 \epsilon -15)+\zeta_{0} (13-15 \epsilon ))+2 \zeta_{0} \zeta (\zeta_{0} (13 \epsilon -11) \\
&-4 \bar{\zeta} (5 \epsilon -4))+2 (2 \bar{\zeta}-\zeta_{0}) \zeta_{0}^2 (5 \epsilon -4)\big) \,, \\
p_{0,2,0} &= \bar{\zeta} \big(\zeta_{0} \left(\bar{\zeta}^2 (15 \epsilon -13)+2 \zeta_{0} \bar{\zeta} (11-13 \epsilon )+2 \zeta_{0}^2 (5 \epsilon -4)\right) \\
&+\zeta \left(\bar{\zeta}^2 (15-19 \epsilon )+8 \zeta_{0} \bar{\zeta} (5 \epsilon -4)+4 \zeta_{0}^2 (4-5 \epsilon )\right)\big) \,, \\
p_{0,0,2} &= (\epsilon -1) (\zeta-\bar{\zeta}) \zeta_{0}^3 \,, \\
p_{1,1,0} &= (\epsilon -1)(\zeta-\bar{\zeta}) \zeta_{0} (2 \zeta_{0} (\zeta_{0}-2 \bar{\zeta})+\zeta (\bar{\zeta}-4 \zeta_{0})) \,, \\
p_{1,0,1} &= (\epsilon -1)\zeta_{0} \big(3 (\bar{\zeta}-2 \zeta_{0}) \zeta^2+\zeta_{0} (10 \zeta_{0}-9 \bar{\zeta}) \zeta \\
&+2 (2 \bar{\zeta}-\zeta_{0}) \zeta_{0}^2\big) \,, \\
p_{0,1,1} &= -(\epsilon -1)\zeta_{0}\big( \zeta \left(3 \bar{\zeta}^2-9 \zeta_{0} \bar{\zeta}+4 \zeta_{0}^2\right) \\
&-2 \zeta_{0} \left(3 \bar{\zeta}^2-5 \zeta_{0} \bar{\zeta}+\zeta_{0}^2\right) \big) \,, \\
p_{1,0,0} &= (3-4 \epsilon ) \big(\zeta^2 (\bar{\zeta} (7 \epsilon -11)+\zeta_{0} (10-7 \epsilon ))+\zeta_{0} \zeta (\bar{\zeta} (19-8 \epsilon ) \\
&+2 \zeta_{0} (4 \epsilon -7))+4 \zeta_{0}^2 (\zeta_{0}-2 \bar{\zeta})\big) \,, \\
p_{0,1,0} &= (4 \epsilon -3) \big(\zeta_{0} \left(\bar{\zeta}^2 (10-7 \epsilon )+2 \zeta_{0} \bar{\zeta} (4 \epsilon -7)+4 \zeta_{0}^2\right) \\
&+\zeta \left(\bar{\zeta}^2 (7 \epsilon -11)+\zeta_{0} \bar{\zeta} (19-8 \epsilon )-8 \zeta_{0}^2\right)\big) \,, \\
p_{0,0,1} &= -\epsilon (4 \epsilon -3) (\zeta-\bar{\zeta}) \zeta_{0}^2 \,, 
\end{align*}
and
\begin{align*}
q_{2,0,0} &= \zeta^2 \big(\zeta (\bar{\zeta} (5-6 \epsilon )+\zeta_{0} (5 \epsilon -3))+\zeta_{0} (\bar{\zeta} (5 \epsilon -4) \\
&+2 \zeta_{0} (1-2 \epsilon ))\big) \,, \\
q_{0,2,0} &= \bar{\zeta}^2 \big(\zeta_{0} (\bar{\zeta} (3-5 \epsilon )+2 \zeta_{0} (2 \epsilon -1))+\zeta (\bar{\zeta} (6 \epsilon -5) \\
&+\zeta_{0} (4-5 \epsilon ))\big) \,, \\
q_{0,0,2} &= (\epsilon -1)(\zeta-\bar{\zeta}) \zeta_{0}^3 \,,\\
q_{1,1,0} &= -4(\epsilon -1) (\zeta-\bar{\zeta})  \zeta \bar{\zeta} \zeta_{0} \,, \\
q_{1,0,1} &= -2 (\epsilon -1) \zeta \zeta_{0} \left(\zeta (\bar{\zeta}+\zeta_{0})-2 \zeta_{0}^2\right) \,, \\
q_{0,1,1} &= 2 (\epsilon -1) \bar{\zeta} \zeta_{0} (\zeta \bar{\zeta}+\zeta_{0} (\bar{\zeta}-2 \zeta_{0})) \,, \\
q_{1,0,0} &=  (4 \epsilon -3) \zeta (\zeta (\bar{\zeta} (3 \epsilon -4)-\zeta_{0} (\epsilon -2))+\zeta_{0} (3 \bar{\zeta}-4 \zeta_{0} \epsilon )) \,, \\
q_{0,1,0} &= - (4 \epsilon -3) \bar{\zeta} (\zeta (\bar{\zeta} (3 \epsilon -4)+3 \zeta_{0})-\zeta_{0} (\bar{\zeta} (\epsilon -2)+4 \zeta_{0} \epsilon )) \,, \\
q_{0,0,1} &= -\epsilon  (4 \epsilon -3)(\zeta-\bar{\zeta}) \zeta_{0}^2 \,.
\end{align*}

The full form of \cref{eq:1l_witten_pf}
\begin{align}
\Ann_{\rm PF}
=
\left(\partial_{\zeta}+r_{1}\right)\cdot\left(\partial_{\zeta}+r_{2}\right)\cdot\left(\partial_{\zeta}^{2}+r_{3}\partial_{\zeta}+r_{4}\right)\,,
\end{align}
where,
\begin{align}
 & r_{1}=\frac{1}{\zeta-\overset{\_}{\zeta}}+\frac{2}{\zeta}+\frac{2}{\zeta-1}\,,\nonumber \\
 & r_{2}=\frac{1}{\zeta-\overset{\_}{\zeta}}+\frac{2}{\zeta}-\frac{2\epsilon-1}{\zeta-1}\,,\nonumber \\
 & r_{3}=-\frac{2(\epsilon-1)}{\zeta-\overset{\_}{\zeta}}-\frac{2\epsilon-1}{\zeta-1}+\frac{2\epsilon}{\zeta}\,,\nonumber \\
 & r_{4}=-\frac{2(\epsilon-1)\epsilon}{\zeta\left(\zeta-\overset{\_}{\zeta}\right)}+\frac{(\epsilon-1)(3\epsilon-1)}{(\zeta-1)\left(\zeta-\overset{\_}{\zeta}\right)}\nonumber \\
 & \qquad+\frac{(\epsilon-1)\epsilon}{\zeta^{2}}-\frac{(2\epsilon-1)\epsilon}{\zeta-1}+\frac{(2\epsilon-1)\epsilon}{\zeta}\,.
\end{align}

\bibliography{refs.bib}

\end{document}